\documentclass[12pt]{iopart}
\usepackage{iopams}
\usepackage{epsfig}
\newcommand{\be}{\begin{equation}}
\newcommand{\ee}{\end{equation}}
\newcommand{\bd}{\begin{displaymath}}
\newcommand{\ed}{\end{displaymath}}
\newcommand{\BE}{\begin{eqnarray}}
\newcommand{\EE}{\end{eqnarray}}

\newcommand{\bra}{\left\langle}
\newcommand{\ket}{\right\rangle}

\newcommand{\bx}{\ensuremath{\mathbf{x}}}

\newcommand{\bC}{\ensuremath{\mathbf{C}}}

\newcommand{\bG}{\ensuremath{\mathbf{G}}}
\newcommand{\bGz}{\ensuremath{\mathbf{G_0}}}

\newcommand{\bJ}{\ensuremath{\mathbf{J}}}

\newcommand{\bM}{\ensuremath{\mathbf{M}}}

\newcommand{\boldpsi}{{\mbox{\boldmath $\psi$}}}

\newcommand{\wt}{\widetilde}
\newcommand{\avg}[1]{\left\langle{#1}\right\rangle}
\newcommand{\davg}[1]{\left\langle\left\langle{#1}\right\rangle\right\rangle}
\begin{document}

\title[Random replicators with asymmetric couplings]{Random replicators with asymmetric couplings}
\author{Tobias Galla\dag\ddag
}

\address{\dag\ The Abdus Salam International Centre for Theoretical Physics, Strada Costiera 11, 34014 Trieste, Italy}
\address{\ddag\ CNR-INFM, Trieste-SISSA Unit, V. Beirut 2-4, 34014 Trieste, Italy}

\begin{abstract}
Systems of interacting random replicators are studied using generating
functional techniques.  While replica analyses of such models are
limited to systems with symmetric couplings, dynamical approaches as
presented here allow specifically to address cases with asymmetric
interactions where there is no Lyapunov function governing the
dynamics. We here focus on replicator models with Gaussian couplings of
general symmetry between $p\geq 2$ species, and discuss how an
effective description of the dynamics can be derived in terms of a
single-species process. Upon making a fixed point ansatz persistent
order parameters in the ergodic stationary states can be extracted
from this process, and different types of phase transitions can be
identified and related to each other. We discuss the effects of
asymmetry in the couplings on the order parameters and the phase
behaviour for $p=2$ and $p=3$. Numerical simulations verify our
theory. For the case of cubic interactions numerical experiments
indicate regimes in which only a finite number of species survives,
even when the thermodynamic limit is considered.
\end{abstract}

\ead{\tt galla@ictp.trieste.it}

\section{Introduction}

Replicator equations describe the evolution of self-reproducing
interacting species within a given framework of limited
resources. They have found wide applications in a variety of fields
including evolutionary game theory, socio-biology, pre-biotic
evolution, optimization theory and population dynamics, where
equivalences to Lotka-Volterra equations can be established
\cite{Book4,Book5,Book6}. So-called random replicator models (RRM) are
also interesting from the point of view of statistical mechanics, as
they constitute complex disordered systems with non-trivial
co-operative behaviour such as phase transitions, ergodicity breaking
and memory effects.

Statistical physics can contribute to the analysis of large systems of
replicators with quenched interactions. Stochastic couplings drawn at random
from some fixed probability distribution here reflect a lack of
knowledge of the detailed interactions of real replicator systems. The
tools also used for the analysis of disordered systems in the physics
context are then well suited to compute typical properties of such
random replicator models, i.e.  averages of observables over the
distribution of couplings.

The first replicator system with quenched random couplings was
introduced and studied with path integral methods by Diederich and
Opper in \cite{DO,OD,OD2}. Most subsequent studies in the statistical
physics community were then based on either replica theory or purely
on computer simulations \cite{OF1,OF2,OF3,O,SF,Tokita, ParisiBiscari,TokitaYasumoti,ChaTok}.

The restriction to replica methods severely limits the types of models
that can be addressed, as static analyses require the existence of a
global Lypunov (or energy) function which is extremised by the
replicator dynamics. This in turn requires the interaction matrix of
the replicators (for pairwise interaction denoted by $J_{ij}$) to be
symmetric against permutations of the interacting species,
$J_{ij}=J_{ji}$ (generalisation to higher-order interactions is
straightforward and will be discussed below). This appears unrealistic
both in the game-theoretic and the biological context.

In the framework of evolutionary game theory replicator equations
describe the evolution of one or several populations of players, who
engage in a game which is played repeatedly \cite{Book6}. The
interaction matrices here encode the payoffs given to the individual
players, these then reproduce according to their success. In the
simplest case of two players $i$ and $j$ participating in the game
$J_{ij}$ would denote the payoff of player $i$ when playing against
$j$, and $J_{ji}$ the corresponding payoff for $j$. Replica theory can
here be applied to the resulting replicator model only when
$J_{ij}=J_{ji}$ for all pairs $(i,j)$ of players. This corresponds to
equal payoffs for both players, which does not appear to be the
typical situation in real games. On the contrary one would like to
introduce an anti-correlation between $J_{ij}$ and $J_{ji}$ so that
player $i$'s payoff is positive, whenever $j$'s is negative 
and vice versa. In particular zero-sum games correspond to the case
$J_{ij}=-J_{ji}$, i.e. full anti-correlation.  Similarly, the
$\{J_{ij}\}$ in biological applications describe interactions between
different species. The fully symmetric case here corresponds to either
reciprocal stimulation or reciprocal inhibition, which again is
unrealistic, as prey-predator relations between two species $i$ and
$j$ would for example require $J_{ij}$ and $J_{ji}$ to be of opposite
signs.

In order to address replicator systems with fully or partially
asymmetric couplings methods other than replica theory are required.
The only existing analytical studies appear to here be the early
papers by Rieger \cite{Rieger} and by Opper and Diederich \cite{OD,OD2},
where RRMs were studied using generating functional techniques.

Such dynamical methods are not concerned with the extremisation of a
Lyapunov function, but address the temporal evolution of the system
directly. Generating functionals therefore provide a powerful tool to
deal with non-equilibrium disordered systems lacking detailed
balance. While they were originally developed to study magnetic
phenomena in spin-glasses \cite{dD,MPV}, they have recently also been
applied to models of interacting agents for example in the so-called
Minority Game \cite{Book1,Book2,Book3}.

The aim of the present paper is to discuss dynamical methods in the
context of random replicator models and to extend the work of
\cite{OD,OD2} (which is concerned with pairwise Gaussian interactions)
to more general Gaussian replicator systems.  To this end we work out
the effective description of the dynamics of such RRMs in terms of a
single-species process, and then extract the relevant order parameters
in the resulting fixed point states, based on an ergodicity
assumption. The breakdown of ergodicity in turn signals the onset of
memory effects, and we compute the corresponding phase diagrams. Some
differences in the types of transitions are here observed depending on
the symmetry of the couplings and the order of the interactions
(e.g. pairwise versus higher-order). Our study is in direct
extension also of \cite{OF1}, where the case of higher-order symmetric
couplings with Gaussian distribution was addressed using replica
theory.

\section{Model definitions}

We consider replicator systems of $N$ interacting species labelled by
Roman indices $i=1,\dots,N$ with time-dependent concentrations
$x_i(t)\geq 0$. Replicator equations are generally of the form
\be\label{eq:rgen}
\frac{d}{dt}x_i(t)=x_i(t)\left(f_i[\bx(t)]-f[\bx(t)]\right),
\ee
with $f_i[\bx]$ the `fitness' or `success' of species $i$ (which
may depend on the concentrations $\bx=(x_1,\dots,x_N)$ of all
species). $f[\bx]$ in turn denotes the `mean' fitness $f[\bx]=\sum_i
x_i f_i[\bx]$. Thus the concentration of species who are fitter than
the average increases, and the weight of species less fit than average
decreases. Note that Eq. (\ref{eq:rgen}) conserves the total
concentration $\sum_i x_i(t)=1$.

We will here focus on systems with second and higher-order
interactions, with couplings drawn from a Gaussian distribution, i.e.
we will consider replicator equations of the form
\be\label{eq:gaussian} \fl
\frac{d}{dt}x_i(t)=-x_i(t)\left[2ux_i(t)+\sum_{(i_2,\dots,i_{p})\in
    M^{(p)}_i}J^{\,i}_{i_2,i_3,\dots,i_{p}}x_{i_2}(t)x_{i_3}(t)\cdots
  x_{i_{p}}(t)-\kappa(t)-h(t)\right], \ee with $p$ a fixed integer and
where $M^{(p)}_i=\{(i_2,\dots,i_{p}):1\leq i_2<i_3<\dots<i_{p}\leq N;
i_2,\dots,i_{p}\neq i\}$. Note the overall negative sign introduced for later convenience.

The first (diagonal) term in the square brackets describes a
self-interaction. Such terms were first introduced in the context of
replicator models in \cite{Book5}. Depending on the value of the model
parameter $u\geq 0$, the so-called `co-operation pressure', the growth
of any single individual concentration is either strongly or only
moderately suppressed (large versus small values of $u$
respectively). Given the fixed total concentration of species large
values of $u$ hence drive the system towards a co-operative state in
which many species survive at small individual concentrations, whereas
small co-operation pressure can lead to stationary states in which a
large fraction of the species dies out asymptotically and where the
remaining ones operate at high concentrations. In terms of game theory
this corresponds to states in which many or few pure strategies are
being played respectively \cite{Book6}. In the limit $u\to\infty$ the
only stable attractor is given by a fixed point at which all species
carry equal weight.

The second term in the square bracket of (\ref{eq:gaussian}) describes
the interaction between species. The couplings $\{J\}$ are assumed to
be drawn from a Gaussian distribution with zero mean and with variance
and correlations as follows \cite{Ritort}:  \be
\overline{(J^{\,i_1}_{i_2,\dots,i_p})^2}=\frac{p!}{2N^{p-1}},
~~~~~~\overline{J^{\,i_1}_{i_2,\dots,i_p}J^{i_k}_{i_1,\dots,i_{k-1},i_{k+1},\dots,i_p}}=\Gamma\frac{p!}{2N^{p-1}}.
\ee 
$\overline{\cdots}$ denotes an average over the distribution of the
couplings. $\Gamma\in[-1,1]$ is a symmetry parameter. For $\Gamma=1$
their distribution is fully symmetric with respect to permutations of
the indices, $\Gamma=0$ corresponds to the case of fully uncorrelated
couplings, and $\Gamma=-1$ to the fully anti-correlated case. Choosing
$-1\leq\Gamma\leq 1$ allows one to interpolate smoothly between the
different regimes. In the case of pairwise interactions $p=2$ the
choice $\Gamma=-1$ corresponds to $J_{ij}=-J_{ji}$, i.e. to zero sum
games as mentioned in the introduction. For $p>2$ the value
$\Gamma=-1$ cannot be realised, since e.g. for $p=3$ it is impossible
to draw three couplings all being equal in their absolute values but
with pairwise opposite signs.

The field $h(t)$ in Eq. (\ref{eq:gaussian}) has been introduced to
generate response functions in the course of the dynamical analysis of
the system. It will be set to zero at the end of the calculation.

Finally, up to a sign the normalisation parameter $\kappa(t)$
corresponds to the mean fitness $f[\bx]$ mentioned above and is chosen
to ensure the constraint
\be\label{eq:norm}
\frac{1}{N}\sum_i x_i(t)=1
\ee
at any time $t$. We choose initial conditions such that this constraint
is fulfilled at the starting point $t=0$. Note that we have here
re-scaled the concentrations such that $\sum_i x_i=N$ instead of
$\sum_i x_i=1$; this is to ensure appropriate scaling of the
statistical mechanics quantities and a well-defined thermodynamic
limit.
\section{Generating functional analysis and fixed point solutions}
\subsection{Effective theory} 
Eqs. (\ref{eq:gaussian}) define a set of $N$ evolution equations, which
are coupled to each other through the random interactions $\bJ$, and
through the overall constraint (\ref{eq:norm}). A Lyapunov
function governing this dynamics can be found only in the
case of symmetric couplings $\Gamma=1$. In the following we will apply
generating functionals originally developed by De Dominicis \cite{dD}
to obtain a macroscopic description of the coupled $N$-species dynamics.

To this end one defines a dynamical generating functional
\BE
Z[\boldpsi]&=&\davg{\exp\left(i\sum_i\int dt \psi_i(t) x_i(t)\right)} \nonumber \\
&=&\int D\bx\, p_0(\bx(0))\exp\left(i\sum_i\int dt \psi_i(t) x_i(t)\right)\prod_{i,t}\delta[\mbox{Equation (\ref{eq:gaussian})}].
\EE
$\davg{\dots}$ denotes an average over all possible paths of the
system. $p_0(\bx(0))$ describes the (possibly random) initial
conditions from which the dynamics is started. All initial
concentrations are taken to be strictly positive with probability one
here, and we assume in the following that the distribution of initial
conditions factorizes over individual species, i.e.
$p_0(\bx(0))=\prod_i p_0(x_i(0))$.

The general procedure then consists in a computation of the
disorder-averaged generating functional $\overline{Z[\boldpsi]}$, from
which dynamical order parameters can be computed as derivatives with
respect to the fields $\{\psi_i(t), h(t)\}$. The aim is to
formulate a closed set of equations describing the temporal evolution
of a suitable set of dynamical observables, which adequately describe
the macroscopic dynamics. The relevant order parameters for
the RRMs considered here are given by the Lagrange parameter
$\bkappa=\{\overline{\avg{\kappa(t)}}\}$ and the correlation and
response functions $\{\bC, \bG\}$ in the thermodynamic limit:
\be\label{eq:cgor}
C(t,t')=\lim_{N\to\infty}\frac{1}{N}\sum_{i=1}^N \overline{\avg{x_i(t)x_i(t')}}, ~~~ G(t,t')=\lim_{N\to\infty}\frac{1}{N}\sum_{i=1}^N \overline{\avg{\frac{\delta x_i(t)}{\delta h(t')}}}. 
\ee
Here $\avg{\dots}$ stands for an average over realisations of the
possibly randomly drawn initial conditions. Note
that $G(t,t')=0$ for $t\leq t'$ due to causality.

The evaluation of the disorder-average in the generating functional
can be performed along the lines of \cite{Crisanti} after a 
transformation $y_i(t)=\log x_i(t)$ has been performed. The
calculation is straightforward, but lengthy. We will therefore not
enter the details of the mathematical derivation here, but will only
report the final outcome.

One finds that a description of the dynamics can be
obtained in terms of an effective process of the form
\be\label{eq:effprocess}
\fl\frac{d}{dt}x(t)=-x(t)\left[2ux(t)-\Gamma\frac{p(p-1)}{2}\int_0^t
  dt' G(t,t')C(t,t')^{p-2}x(t')+\eta(t)-\kappa(t)-h(t)\right]. 
\ee
This equation describes a stochastic process for a representative
species concentration $\{x(t)\}$, which is a random variable and subject
to Gaussian noise $\{\eta(t)\}$, with non-trivial temporal
correlations according to
\be\label{eq:noise}
\bra \eta(t)\eta(t')\ket_\star=\frac{p}{2}C(t,t')^{p-1}.
\ee

The correlation and response functions $\bC$ and $\bG$ of the original
multi-species system (see Eq. (\ref{eq:cgor})) can in turn be shown to
be given by \be\label{eq:selfcons} C(t,t')=\avg{x(t)x(t')}_\star,~~~
G(t,t')=\avg{\frac{\delta x(t)}{\delta h(t')}}_\star, \ee 
and $\bkappa$ in (\ref{eq:effprocess}) 
has to be chosen such that 
\be\label{eq:normeff} \avg{x(t)}_\star = 1 \qquad \forall t.  
\ee 

$\avg{\dots}_\star$ denotes an average over realisations of the
effective process (\ref{eq:effprocess}), i.e.  over realisations of
the single-species coloured noise $\{\eta(t)\}$ and over initial
conditions described by $p_0(x(0))$. 

Eqs.
(\ref{eq:effprocess},\ref{eq:noise},\ref{eq:selfcons},\ref{eq:normeff})
thus form an implicit, but closed set of equations from which
$\{\bkappa,\bC,\bG\}$ are to be obtained. For $p=2$ this system
coincides with the results of \cite{OD}. The description in terms of
an ensemble of decoupled, but stochastic effective species is exact
and fully equivalent to the original $N$-species problem (defined by
Eq. (\ref{eq:gaussian})) in the limit $N\to\infty$ (in the sense that
disorder-averages of macroscopic observables in the original dynamics
can be obtained as averages $\avg{\dots}_\star$ on the level of the
effective process). The retarded self-interaction and the coloured
noise in the effective process are direct consequences of the quenched
disorder in the original problem and impede a full explicit analytical
solution of the self-consistent saddle-point problem for the two-time
quantities $C(t,t')$ and $G(t,t')$ and the full function
$\kappa(t)$. One is therefore limited to a specific ansatz for the
trajectories of the effective particles.

\subsection{Fixed point solutions}
In order to proceed analytically we assume that the system reaches a
fixed point asymptotically, i.e. $x_i(t)\to x_i$ as $t\to\infty$ for
all $i$ in the original dynamics. Numerical simulations confirm that
this assumption is indeed justified for values of $u$ larger than some
critical value $u_c$ (the value of $u_c$ may depend on $p$ and $\Gamma$ as discussed in detail below).

We may then make a similar assumption for the realisations of the
effective process, i.e. $x(t)\to x$, with $x$ a static random
variable. 
Accordingly the correlation function becomes flat at the
fixed point and $\kappa(t)$ approaches a fixed point as well, so that we write \be \lim_{t\to\infty} C(t+\tau,t)=q~~~\forall \tau, ~~~
\lim_{t\to\infty}\kappa(t)=\kappa. \ee We also make the standard ansatz
of a time-translation invariant asymptotic response function 
\be
\lim_{t\to\infty} G(t+\tau,t)=G(\tau).  
\ee 

Furthermore we will only address
ergodic stationary states, that is states in which perturbations have
no long-term effects so that the integrated response function remains
finite 
\be \chi\equiv\int_0^\infty d\tau\,
G(\tau)<\infty, \ee 
and no long-term memory is present, i.e. we will
assume 
\be \lim_{t\to\infty} G(t,t')=0 \ee 
also for finite $t'$. In the absence of memory we may send the starting point of the dynamics
to $-\infty$ for convenience.

\begin{figure}[t!]
\begin{center}
\includegraphics[width=6.5cm]{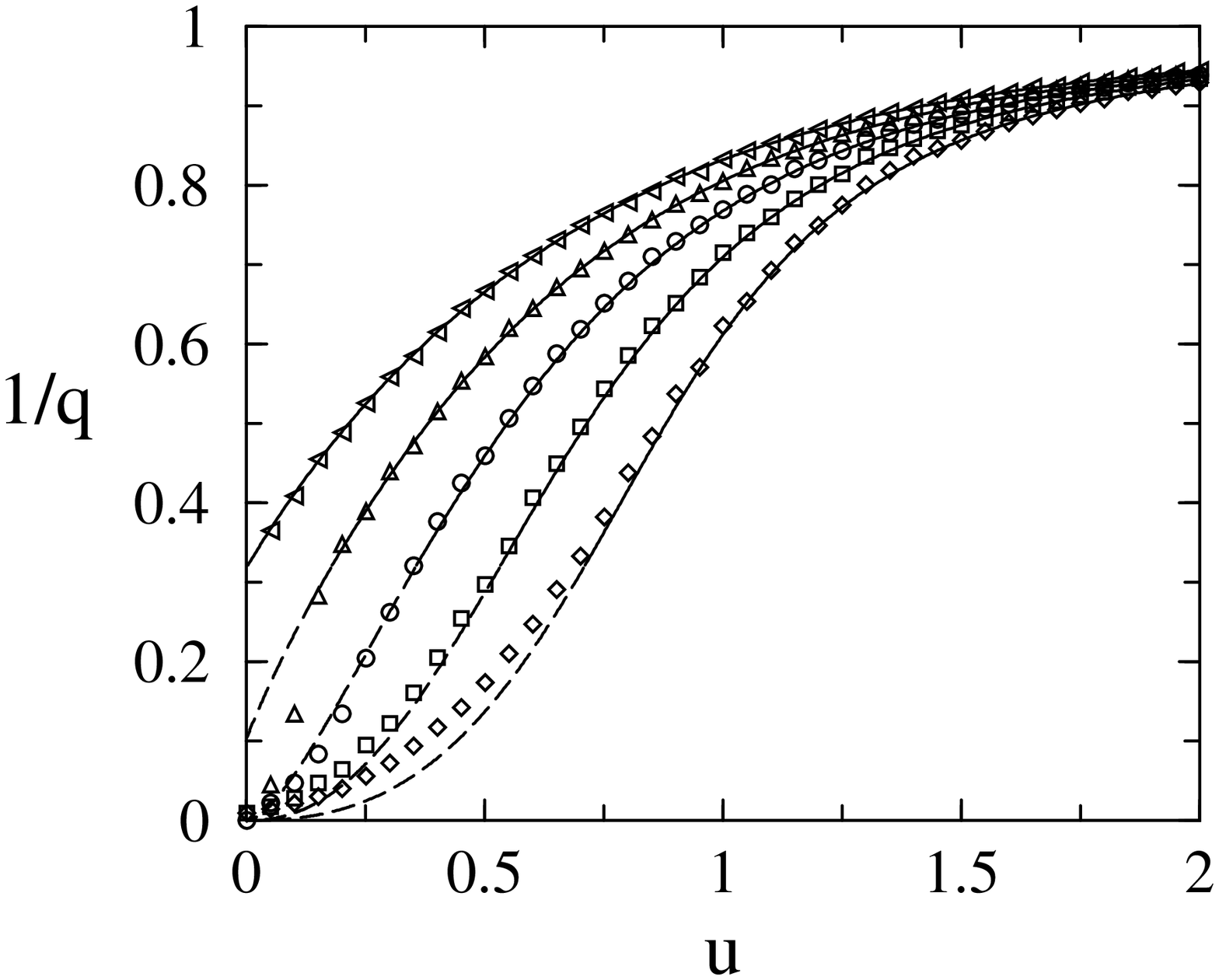}~~~~~
\includegraphics[width=6.5cm]{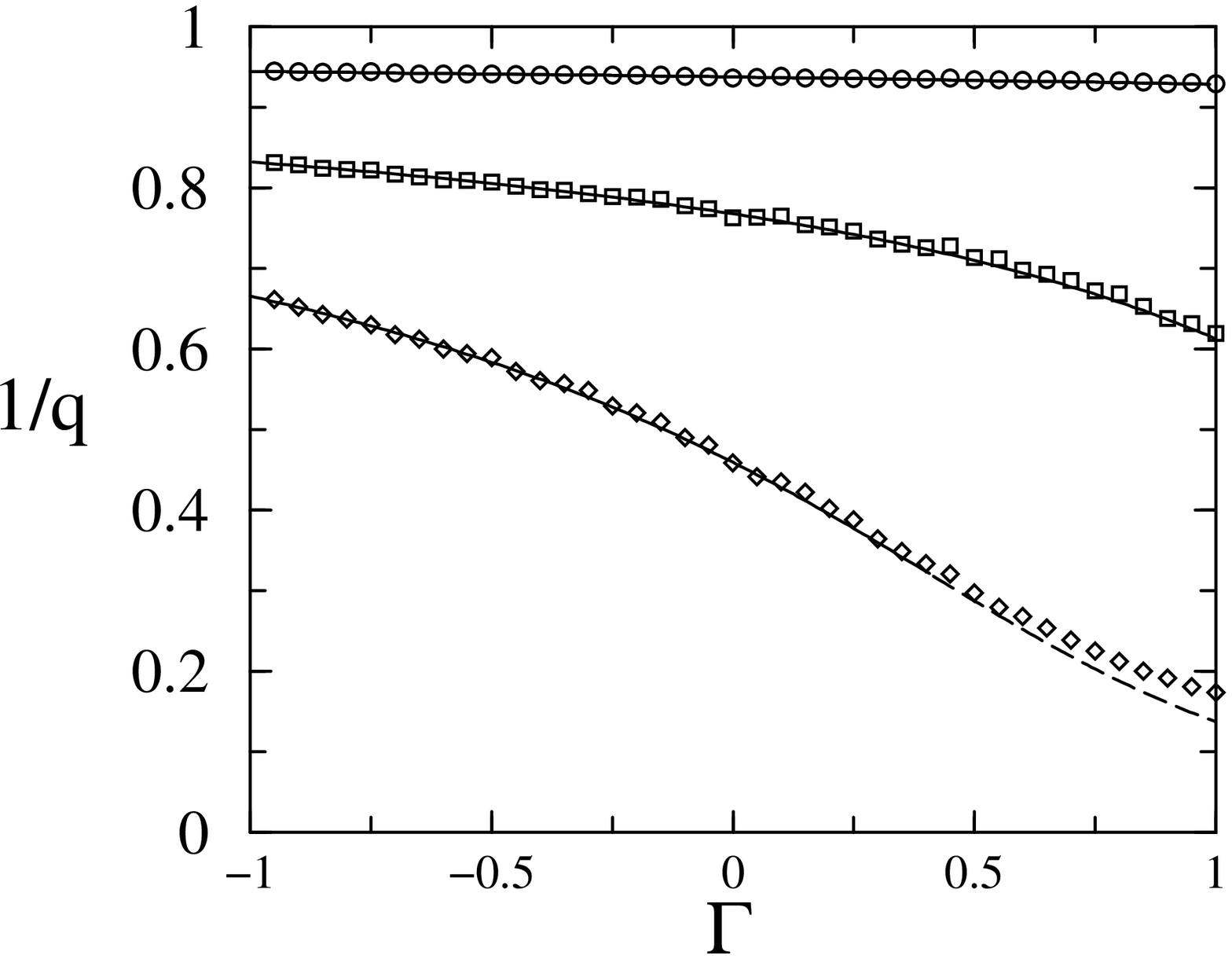}\\
\end{center}
\caption{\label{fig:p_2qphi} Inverse order parameter $q$ for the model
  with $p=2$. Symbols are from simulations for $N=300$ species, run
  for $20000$ discretisation steps (effective time-stepping is of
  order $0.01-0.1$) and averaged over $50$ samples of the
  disorder. Measurements are taken in the last quarter of the
  simulations. Initial conditions correspond to a flat distribution
  over the interval $[0,2]$.  The solid lines represent the analytical
  predictions in the ergodic phase, and have been continued as dashed
  lines into the region of broken ergodicity, where our theory is no
  longer valid. {\bf Left:} $1/q$ as a function of $u$ for
  $\Gamma=-1,-0.5,0,0.5,1$ from top to bottom. {\bf Right:} Same
  quantity as a function of $\Gamma$ at fixed $u=2,1,0.5$ (top to
  bottom).}
\end{figure}

Finally within the fixed-point ansatz we also assume that each
realisation of the the single-species noise $\{\eta(t)\}$ approaches a
time-independent value $\eta$ asymptotically, which according to
(\ref{eq:noise}) is then a static Gaussian variable with zero mean and
variance
\be
\avg{\eta^2}_\star=\frac{p}{2}q^{p-1}. 
\ee
Fixed points of (\ref{eq:effprocess}) then fulfill the condition
\be
x\left(2ux-\Gamma\frac{p(p-1)}{2}q^{p-2}\chi
  x+\left(\frac{p}{2}q^{p-1}\right)^{1/2}z-\kappa\right)=0,
\ee
where we have written $\eta=\left(\frac{p}{2}q^{p-1}\right)^{1/2}z$
with $z$ a standard Gaussian variable. $h(t)$ has been set to zero at
this stage (derivatives with respect to $h(t)$ can be obtained as
derivatives with respect to the effective-particle noise $\eta(t)$ up
to an inverted sign). We note that $x(z)=0$ is a solution of this
equation for all realisations of the random variable $z$. Other
solutions may be possible whenever setting the bracket to zero leads
to a positive value of $x(z)$. Taking into account the stability of
the zero solutions, one finds that the ansatz
\be\label{eq:fp}
x(z)=\frac{\kappa-\left(\frac{p}{2}q^{p-1}\right)^{1/2}z}{2u-\Gamma p(p-1)\chi q^{p-2}/2}\Theta\left[\kappa-\left(\frac{p}{2}q^{p-1}\right)^{1/2}z\right].
\ee
adequately describes the solutions, with $\Theta[x]$ the step function ($\Theta[x]=1$ for $x>0$ and $\Theta[x]=0$ otherwise). Similarly to
\cite{OD,myreplicator} one demonstrates that zero fixed points $x(z)=0$ are
unstable for positive arguments of the $\Theta$-function.
Self-consistency then demands that
\BE\label{eq:sp0}
\fl~~~\int Dz~ x(z)=1, ~~~ \int Dz~ x(z)^2 =q, ~~~~-\left(\frac{p}{2}q^{p-1}\right)^{-1/2}\int Dz\frac{\partial
  x(z)}{\partial z}=\chi,
\EE
with $Dz=\frac{dz}{\sqrt{2\pi}}e^{-z^2/2}$ a standard Gaussian
measure.
Upon using the explicit ansatz (\ref{eq:fp}) for the fixed points these equations may be written as
\BE
\left(\frac{p}{2}q^{p-1}\right)^{-1/2}\left(2u-\Gamma\frac{p(p-1)}{2}\chi q^{p-2}\right)&=&\int_{-\infty}^\Delta Dz (\Delta-z), \label{eq:sp1}\\
q\left(\frac{p}{2}q^{p-1}\right)^{-1}\left(2u-\Gamma\frac{p(p-1)}{2}\chi q^{p-2}\right)^2&=&\int_{-\infty}^\Delta Dz (\Delta-z)^2, \label{eq:sp2} \\
\left(2u-\Gamma\frac{p(p-1)}{2}\chi q^{p-2}\right)\chi&=&\int_{-\infty}^\Delta Dz \label{eq:sp3},
\EE
with $\Delta=\kappa/\left(\frac{p}{2}q^{p-1}\right)^{1/2}$.  We note that $\phi=\int_{-\infty}^\Delta Dz$ represents the
probability of a given species $i$ to survive in the long-term limit,
i.e. to attain a fixed point value $x_i=\lim_{t\to\infty}x_i(t)>0$. We
will refer to $\phi$ as the fraction of surviving species in the
following. For the case of symmetric couplings, $\Gamma=1$, equations (\ref{eq:sp1}, \ref{eq:sp2},
\ref{eq:sp3}) are identical to those found in replica-symmetric
studies of the statics of the model \cite{OF1}, for $p=2$ they furthermore 
coincide with the ones reported in \cite{OD}. They are readily solved
numerically (in terms of $\kappa,\, q$ and $\chi$) for arbitrary
values of the model parameters $u$ and $\Gamma$ (at fixed
$p$)\footnote{Note here that the right-hand-sides of (\ref{eq:sp1},
\ref{eq:sp2}, \ref{eq:sp3}) can be written in closed form as functions
of $\Delta$ after performing the Gaussian integrals over $z$. A
solution of these equations can thus most efficiently be obtained by
expressing $\{u,q,\chi\}$ as functions of $\Delta$, and by
subsequently varying $\Delta$ \cite{SF}.}. Results for the order
parameter $q$ are depicted for the model with pairwise interaction
($p=2$) in Fig. \ref{fig:p_2qphi}. We find near perfect agreement
between the analytical theory and numerical simulations\footnote{A
discretisation method similar to that of \cite{OD2} is used.  The
typical (effective) time-step is of the order of $\Delta t\approx
0.01-0.1$.} for $u$ larger than some critical value $u_c(\Gamma)$ (for
$\Gamma<\Gamma_c(u)$ respectively in the right panel of
Fig. \ref{fig:p_2qphi}). The deviations from the theory below $u_c$
(resp. above $\Gamma_c$) are due to a breakdown of some of our
assumptions regarding ergodicity and the stability of the fixed
points, the resulting phase transitions will be discussed in more
detail below.

Let us finally in this section turn to an interpretation of the order
parameter $q$, which within the fixed point ansatz is given by
$q=\lim_{N\to\infty}N^{-1}\sum_i\overline{\avg{x_i^2}}$. If the $x_i$
were normalised to $1$ instead of $N$, $q$ would be closely related to
what is known as Simpson's index of diversity \cite{simpson} and would
indicate the probability that two randomly selected individuals of the
eco-system are of the same species. A Simpson index of zero thus
indicates infinite diversity, a value of one corresponds to no
diversity. A similar interpretation holds for the normalisation of the
$\{x_i\}$ used here: finite values of $q$ indicate the co-existence of
an extensive number of species with concentrations $x_i\sim{\cal
O}(N^0)$, while a divergence in $q$ signals the dominance of a
sub-extensive number of species with diverging concentrations in the
thermodynamic limit \cite{OF2}. $1/q$ is thus a measure for the
diversity of the system, and we expect the behaviour of $1/q$ to be
similar to that of the fraction of surviving species $\phi\equiv\int
Dz~\Theta[x(z)]$. This is indeed confirmed in numerical simulations
(not shown here) in which one verifies that $1/q$ and $\phi$ show the
same qualitative behaviour as functions of the model parameters. The left panel of Fig. \ref{fig:p_2qphi} thus confirms the role of $u$ as a co-operation pressure, the higher the value of $u$ the more species survive in the long-term limit and the more diverse the system is in its stationary state. The influence of the symmetry of the interactions on the diversity (right panel of Fig. \ref{fig:p_2qphi}) will be discussed below.

\subsection{Stability analysis and phase transitions}
Our theory and solution crucially relies on
the assumptions made with respect to ergodicity, weak long-term
memory, and the stability of the fixed point which the dynamics is
assumed to approach asymptotically. A breakdown of this theory may
thus be signalled by the failure of any of these assumptions,
marking a dynamical phase transition at which the ergodic fixed-point regime
ceases to exist. In general one may imagine the following phenomena to
occur:
\begin{enumerate}
\item The analytical ergodic theory resulting in Eqs. (\ref{eq:sp0}) might predict the integrated response $\chi$ to diverge in some subset of the parameter space of the model. Such a transition has been observed in replicator models with Hebbian interactions \cite{myreplicator}, and below this transition the system is found to be non-ergodic.
\item In a similar way the theory might {\em analytically} predict singularities in either $q$ or $\kappa$. No such types of transitions seem to have been observed so far in studies of RRM\footnote{Phases in which $\lim_{N\to\infty}q^{-1}=0$ have been hinted at in \cite{OF1} and will be discussed in more detail below. Note however that in these cases the equations describing the ergodic states do not analytically lead to singularities in $q$, but do not allow for any solutions at all in the phases where $q$ is found to diverge in simulations; they are of the type (iii) in the above list of possible transitions.}.
\item The set of equations  (\ref{eq:sp0}) might fail to have solutions for certain values of the control parameters, so that no ergodic states are allowed in this region of the phase diagram. We will discuss cases of this type in section \ref{sec:gaussian}.
\item Our ergodic theory is based on the assumption that all trajectories of the system will evolve into fixed points asymptotically. An onset of instability of these fixed points against small perturbations will thus signal the breakdown of our theory as they will no longer be local attractors of the dynamics. Such transitions have been observed in RRM with Gaussian and Hebbian couplings in \cite{OD,OD2,myreplicator}. 
\item Finally a breakdown of our requirement that long term memory be weak would indicate that the system remembers perturbations during the transient dynamics. Even if the system still evolves into a fixed point the latter might no longer be unique and the choice of the asymptotic stationary state might thus depend on initial conditions or perturbations at early stages of the temporal evolution. This type of transition has been related to the previous one (instability of the assumed fixed point) in \cite{myreplicator}, and we will demonstrate below that the conditions for both types of transitions are indeed fulfilled at the same points in parameter space whenever a replicator system with symmetric couplings is considered. 
\end{enumerate}

Transitions of the types (i)-(iii) are easily detected in numerical
solutions of Eqs.  (\ref{eq:sp1},\ref{eq:sp2},\ref{eq:sp3}) (or their
analogues for other types of RRMs). In the
following we will therefore focus on (iv) and (v) and will derive
explicit conditions for the onsets of instability and of memory at
finite integrated response respectively.
\subsubsection{Instability of the fixed point}
In order to inspect the stability of the assumed fixed point we will
follow \cite{OD} and add a small component $\varepsilon\zeta(t)$ of white noise
 to the
system (with $\avg{\zeta(t)}=0,\,\avg{\zeta(t)\zeta(t')}=\delta(t-t')$) and then study small fluctuations $\varepsilon
y(t)$ and $\varepsilon v(t)$ of the effective species concentration
and the single-particle noise about their respective fixed points $x$ and $\eta$:
\be\label{eq:perturbedansatz}
x(t)=x+\varepsilon y(t),\qquad \eta(t)=\eta+\varepsilon v(t).
\ee
In the following we will only consider the case $x>0$, as it
turns out that the onset of instability in replicator systems
is determined by the fixed points different from zero while the zero
fixed points remain stable against perturbations (see also
\cite{OD,myreplicator}).
Inserting (\ref{eq:perturbedansatz}) into the effective process
(\ref{eq:effprocess}) (for fixed points $x>0$) and adding the additional noise
component leads to 
\be
\fl\frac{d}{dt}y(t)=-x\left[2uy(t)-\Gamma\frac{p(p-1)}{2}q^{p-2}\int^t_{-\infty} dt' G(t-t') y(t')+v(t)+\zeta(t)\right],
\ee
to first order. In Fourier space one has
\be\label{eq:fou}
\wt y(\omega)=-\frac{\wt v(\omega)+\wt \zeta(\omega)}{\frac{i\omega}{x}+2u-\Gamma\frac{p(p-1)}{2}q^{p-2}\widetilde G(\omega)}.
\ee
(with $\{\widetilde y(\omega),\widetilde v(\omega),\widetilde\zeta(\omega),\widetilde G(\omega)\}$ the Fourier transforms of the one-time functions $\{y(t),v(t),\zeta(t),G(\tau)\}$). Focusing on $\omega=0$ one obtains
\BE\label{eq:dong}
\avg{|\wt y(\omega=0)|^2}_\star=\phi \times\left[\avg{|\wt v(\omega=0)|^2}_\star+1\right]\frac{\chi^2}{\phi^2}.
\EE
The first factor of $\phi$ in (\ref{eq:dong}) takes into account that
(\ref{eq:fou}) was derived for non-zero fixed points $x>0$ and that
fluctuations about zero fixed points do not contribute to $\avg{|\wt
y(\omega=0)|^2}_\star$. We have also used Eq. (\ref{eq:sp3}) to
replace $(2u-\Gamma p(p-1)q^{p-2}\chi/2)^{-2}$ by $\chi^2/\phi^2$. The
self-consistency condition on the covariance of the single-particle
noise (Eq. (\ref{eq:noise})) on the other hand implies
\be\label{eq:stump}
\avg{|\widetilde v(\omega)|^2}_\star=\frac{p(p-1)}{2}q^{p-2}\avg{|\widetilde y(\omega)|^2}_\star
\ee

Insertion of  (\ref{eq:stump}) into (\ref{eq:dong}) then allows one to solve for
$\avg{|\widetilde y(0)|^2}_\star$, and one finds
\be\label{eq:zzzz}
\avg{|\widetilde y(0)|^2}_\star=\left[\frac{\phi}{\chi^2}-\frac{p(p-1)}{2}q^{p-2}\right]^{-1}.
\ee We conclude that $\avg{|\widetilde
y(0)|^2}_\star$ diverges when the square bracket on the
right-hand-side of (\ref{eq:zzzz}) vanishes, suggesting an onset of
instability. $\avg{|\widetilde y(0)|^2}_\star$ is predicted to become
negative whenever the right-hand-side of (\ref{eq:zzzz}) becomes negative, indicating a
further contradiction. The onset of instability thus occurs at the point
defined by
\be\label{eq:instp}
\frac{p(p-1)q^{p-2}}{2\phi}\chi^2=1,
\ee
and the fixed points are stable whenever this
expression is strictly smaller than one. With some further algebra one can show that (\ref{eq:instp}) defines a line $u_c(\Gamma,p)=u_c(\Gamma=0,p)(1+\Gamma)$ in the $(u,\Gamma)$ plane for any fixed value of $p$.

\subsubsection{Memory onset}
In the previous section we have related the breakdown of our
ergodic theory to a local instability of the fixed point reached by
the dynamics. We will now inspect for an onset of long-term memory at
finite integrated response. This type of transition has been observed
previously for example in Minority Games with self-impact correction
or diluted interactions \cite{corr,dilute}, and can also be
interpreted in terms of a breakdown of time-translation invariance, see
\cite{Book2} for details. It is also found in replicator systems with Hebbian interactions \cite{myreplicator}. In order to see how solutions with memory bifurcate from the time-translation invariant ergodic states we will proceed along the lines of \cite{Book2} and make the
following ansatz for the response function
\be\label{eq:g0ghat}
G(t,t')=G_0(t-t')+\varepsilon\widehat G(t,t'),
\ee
where $\varepsilon\widehat G(t,t')$ is a small contribution which
breaks time-translation invariance. The starting point of the dynamics
can here no longer be sent to $-\infty$. For a physical interpretation
of $\bGz$ and $\widehat\bG$ see
\cite{myreplicator}. Following \cite{Book2} $\widehat \bG$ is taken to
depend only on the (earlier) time $t'$ asymptotically,
i.e. $\lim_{t\to\infty}\widehat G(t,t')=\widehat G(t')$.

Starting from (\ref{eq:g0ghat}) one expands the kernel of the retarded self-interaction in the effective process to linear order in $\widehat G$, and finds
the following effective process:
\BE
\hspace{-2.3cm}\frac{d}{dt}x(t)&=&-x(t)\bigg[2ux(t)-\Gamma\frac{p(p-1)}{2}\int^t_0 dt' G_0(t-t')C(t-t')^{p-2} x(t')\\
&&-\varepsilon\Gamma\frac{p(p-1)}{2}\int^t_0 dt' \widehat G(t')C(t-t')^{p-2}x(t')+\eta(t)-\kappa(t)\bigg]\label{eq:effprocessmo},
\EE
so that non-zero asymptotic fixed points are now found to fulfill
\be\label{eq:blieh}
\fl 2ux-\Gamma\frac{p(p-1)}{2}\chi q^{p-2} x -\varepsilon\Gamma\frac{p(p-1)}{2}\int^t_0 ~ dt'\widehat G(t')q^{p-2}x(t')+\left(\frac{p}{2}q^{p-1}\right)^{1/2}z-\kappa=0.
\ee
From this one can compute the response function $\widehat G(t'')$ at a transient time $t''$ self-consistently and finds
\be\label{eq:dadada}
\fl~~~~~\widehat G(t'')=\Gamma\frac{p(p-1)}{2}q^{p-2} \left(2u-\Gamma\frac{p(p-1)}{2}\chi q^{p-2}\right)^{-1}\int_0^t dt' \widehat G(t')\avg{\frac{\delta x(t')}{\delta h(t'')}}_\star.
\ee
This can be understood as an eigenvalue
problem of the form $\widehat G(t)=\int dt' M(t-t')
\widehat G(t')$, with a suitable kernel $\bM$ \cite{Book2}. Upon taking Fourier
transforms and focusing on the zero frequency mode $\omega=0$ one finds after some simpifications
\be
\widehat\chi=\Gamma\frac{p(p-1)}{2\phi}q^{p-2}\chi^2\widehat\chi ,
\ee
with $\widehat\chi=\int dt \widehat G(t)$. Although $\widehat\chi=0$ is always a solution, one realises that non-zero solutions for
$\widehat\chi$ become possible at the point at which
\be\label{eq:mop}
\Gamma\frac{p(p-1)q^{p-2}}{2\phi}\chi^2=1
\ee
We will refer to this as the memory onset (MO) condition in the following.

\section{Phase diagram and discussion of results}\label{sec:gaussian}

\begin{figure}[t!]
\begin{center}
\includegraphics[width=6.5cm]{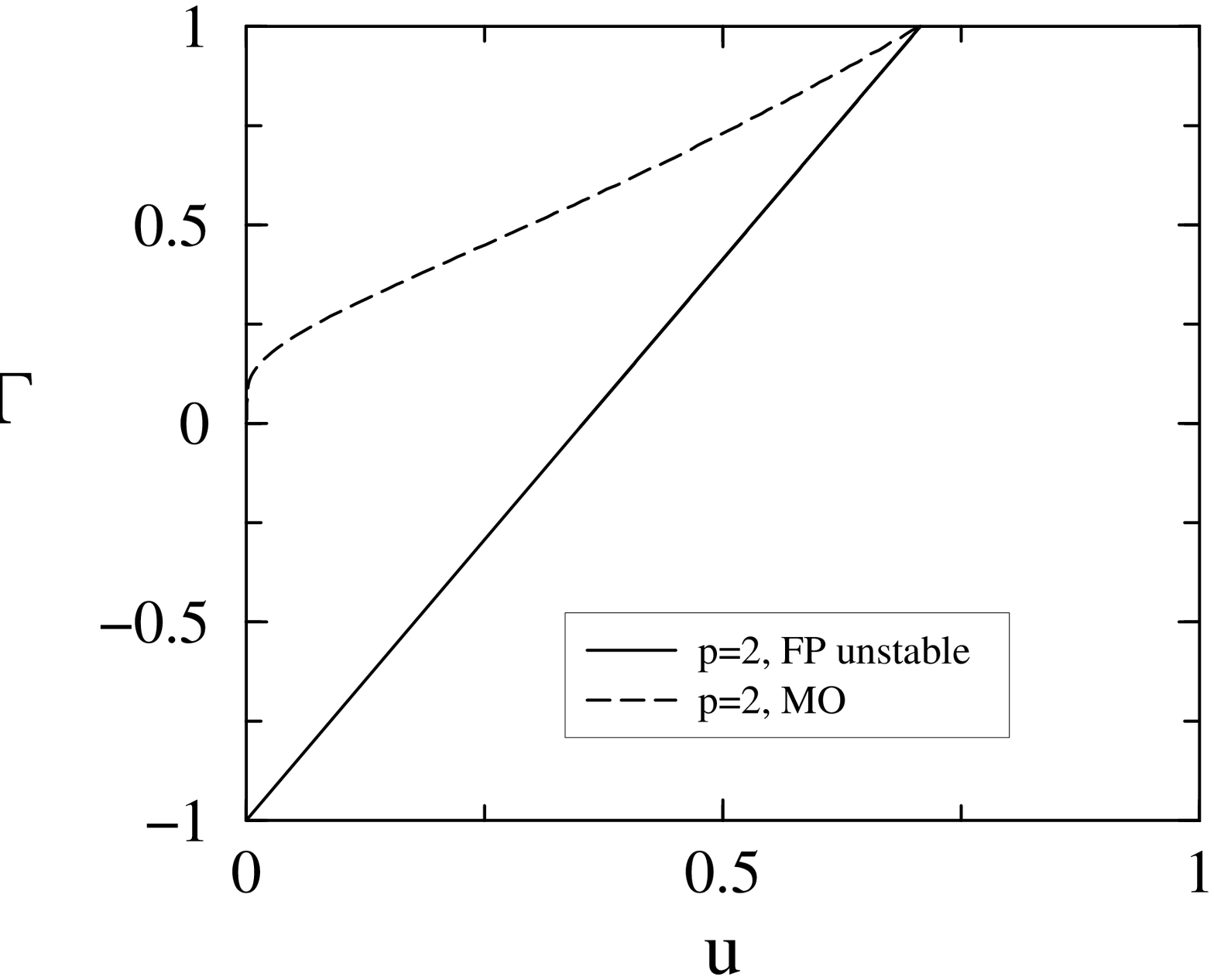}~~~~~
\includegraphics[width=6.5cm]{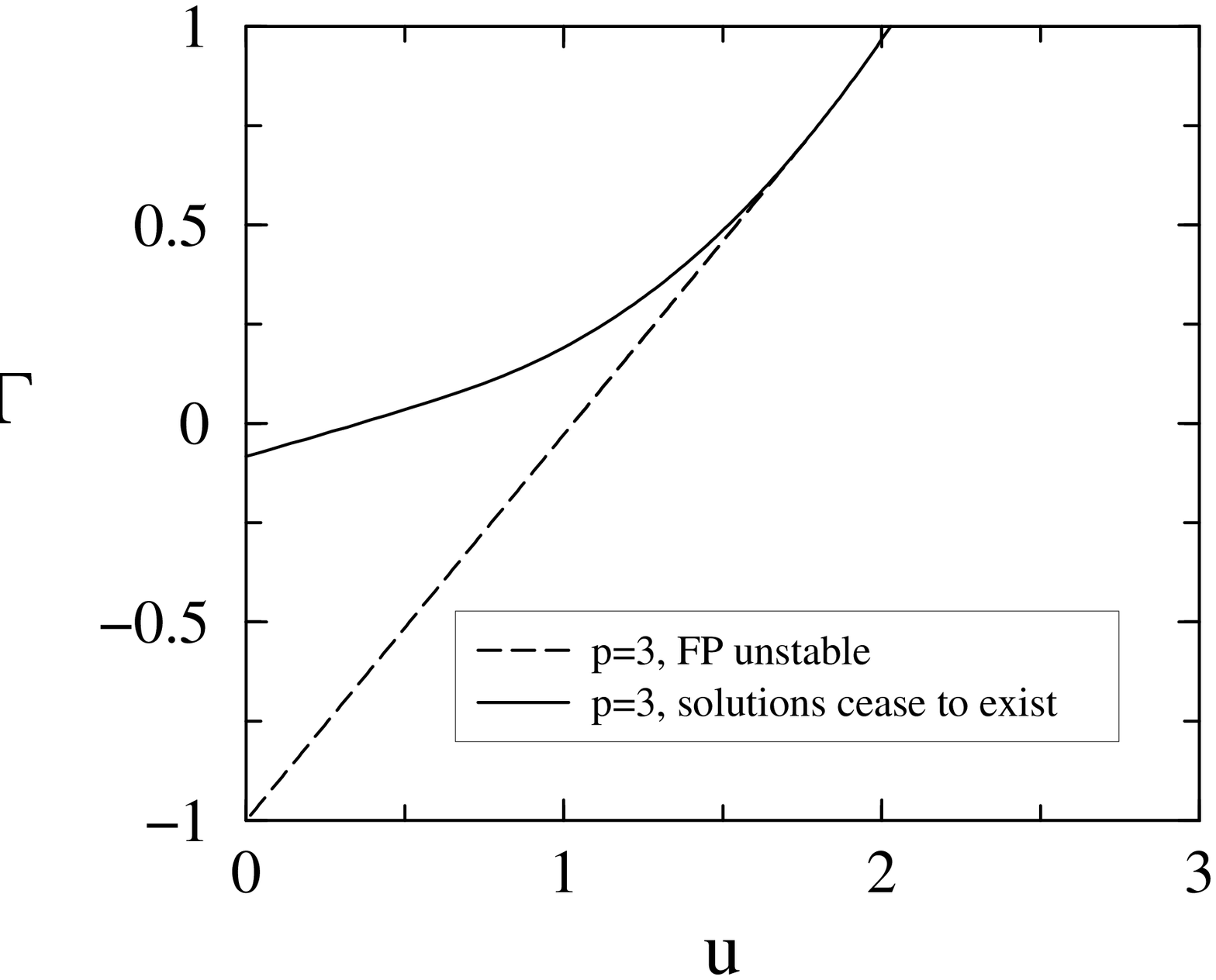}\\
\end{center}
\caption{\label{fig:phasediagram}Phase diagrams for the models wit $p=2$ and $p=3$ in the $(u,\Gamma)$ plane. Left panel ($p=2$): the dashed line marks the onset of long-term memory, see Eq. (\ref{eq:mop}), the solid line marks the onset of the instability of the fixed points, Eq. (\ref{eq:instp}). Right panel ($p=3$): the solid curve separates the region in which Eqs. (\ref{eq:sp1})-(\ref{eq:sp3}) admit a solution (high $u$) from the region in which no solutions are found, dashed line marks the onset of instability, Eq. (\ref{eq:instp}). }
\end{figure}
 
We now turn to a discussion of the resulting phase diagrams, as
depicted in Fig. \ref{fig:phasediagram}. In general we find that our
ergodic fixed-point ansatz is valid and stable for values $u>u_c$ with
$u_c$ some critical value which depends on the order of interaction
$p$ and on the symmetry parameter $\Gamma$. Above $u_c$ the system evolves into a unique fixed point, with virtually all species surviving in the limit $u\to\infty$. This corresponds to a fixed point in the interior of the simplex defined by the condition $N^{-1}\sum_i x_i=1$. For small values of $u$ fewer species survive so that the system operates near the boundary of this simplex.

The dependence of the phase diagram on $\Gamma$ is particularly
interesting. One notes that both conditions (\ref{eq:instp}) and
(\ref{eq:mop}) coincide for symmetric couplings $\Gamma=1$, but that
the onset of the instability of the fixed point occurs before memory
sets in for $\Gamma<1$ \footnote{The equivalence of the MO and
instability conditions at $\Gamma=1$ can be extended to more general
replicator equations derived from a Lyapunov function.}. At the same
time simulations reveal different types of behaviour below the
transition, depending on the symmetry of the interactions: while for
fully symmetric interactions $\Gamma=1$ the system reaches a fixed
point also below $u_c$ , fixed points are not necessarily observed
below $u_c$ for even partially asymmetric couplings. This will be
discussed in more detail below. The type of transition also depends on
whether the interactions are pairwise or cubic, we find that in the
system with $p=2$ the number of surviving species is always extensive
for any $u>0$, whereas this number can turn out to be ${\cal O}(1)$ in
the model with $p=3$ below the transition even in the thermodynamic
limit. In the following we will treat the cases $p=2$ and $p=3$
separately.

\subsection{Pairwise interaction ($p=2$)}
\begin{figure}[t!]
\begin{center}
\includegraphics[width=6.5cm]{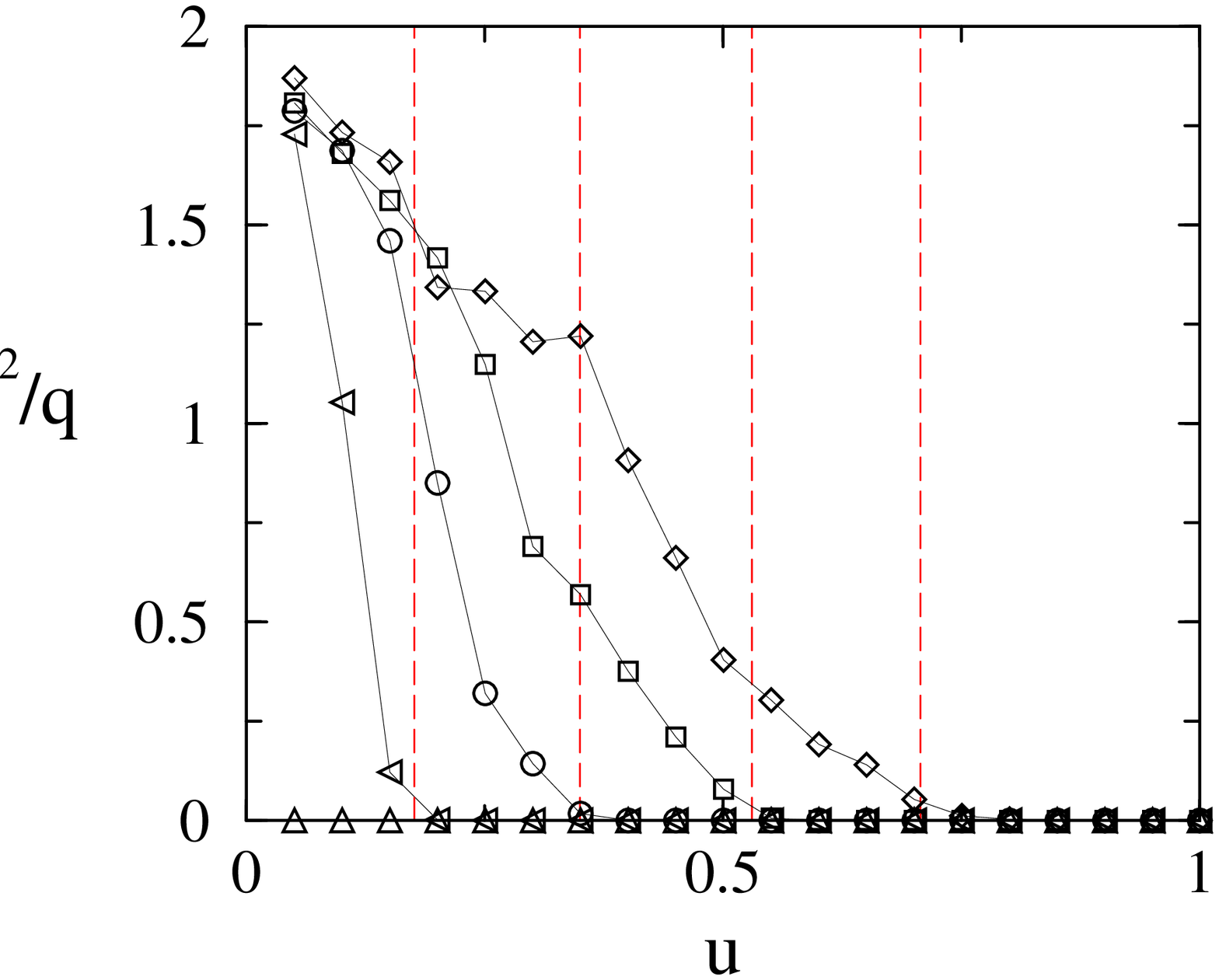}~~~~~
\includegraphics[width=6.5cm]{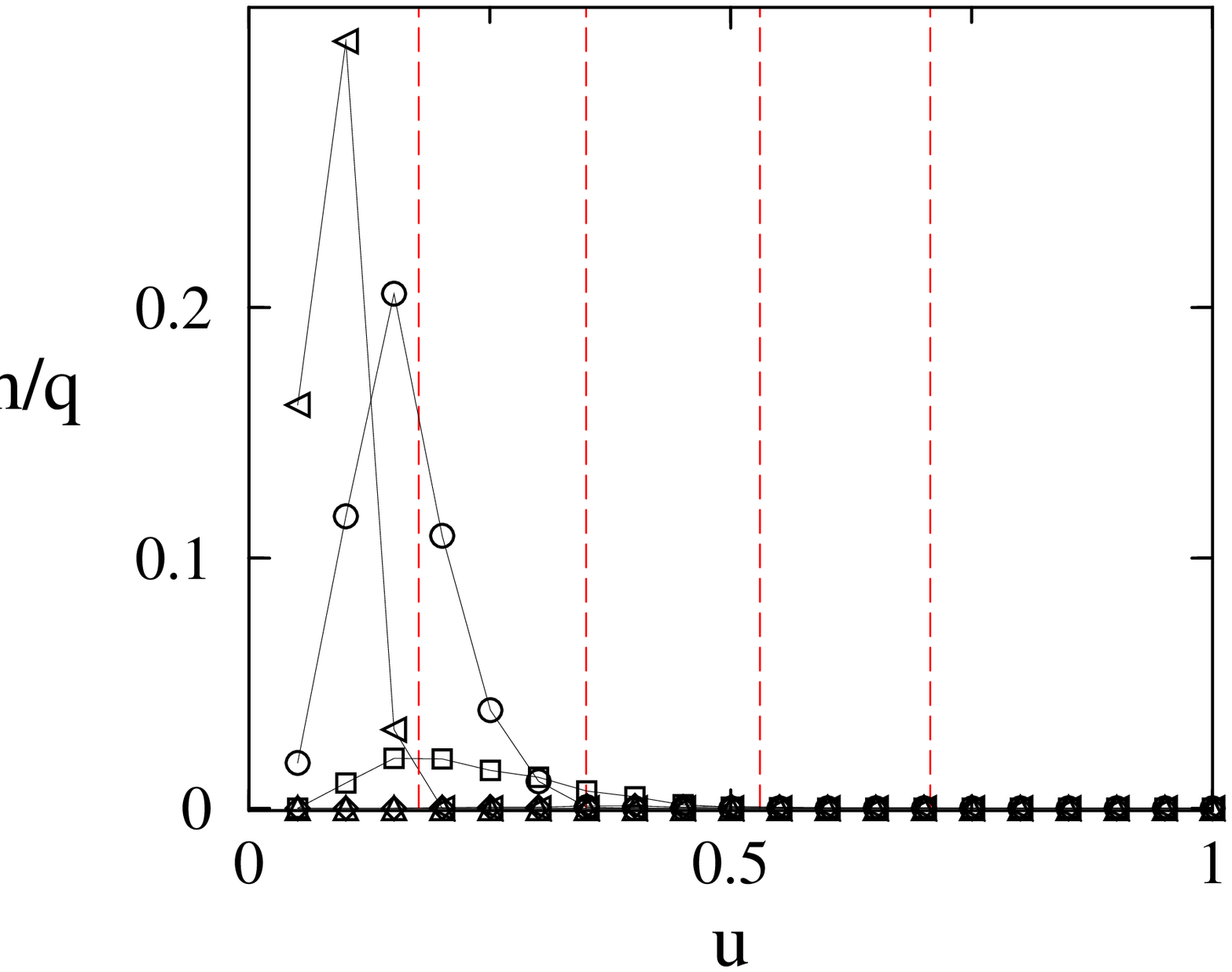}
\caption{\label{fig:p_2_distrough}Relative distance $d^2/q$ and roughness $h/q$ versus $u$ for the model with $p=2$. Connected symbols are from simulations (triangle up: $\Gamma=-1$, triangle left: $\Gamma=-0.5$, circles: $\Gamma=0$, squares: $\Gamma=0.5$, diamonds: $\Gamma=1$). $d^2$ is obtained by comparing the stationary states of two runs at fixed disorder started from independent random initial conditions over $[0,2]$. Simulation parameters as in Fig. \ref{fig:p_2qphi}. Vertical dashed lines indicate the location of the phase transitions as predicted by the theory for $\Gamma=-0.5,0,0.5,1$ from left to right. No transition occurs for $u>0$ and $\Gamma=-1$.}
\end{center}
\end{figure}

We find here that Eqs. (\ref{eq:sp1},\ref{eq:sp2},\ref{eq:sp3}) admit
solutions for all $u$ and $\Gamma$ and we observe no singularities in
any of the persistent order parameters. The predictions of the
analytical theory for the order parameters in the ergodic phase are
compared with numerical simulations in Fig. \ref{fig:p_2qphi} and we
find near perfect agreement in the ergodic phases ($u>u_c(\Gamma)$ and
$\Gamma<\Gamma_c(u)$ respectively).

The only possible violations of our assumptions of the ergodic stationary
state are instabilities of the fixed points or an onset of long-term
memory at finite susceptibility $\chi$. The condition for onset of the
instability of the fixed points (\ref{eq:instp}) reduces to
$u_{c}(p=2,\Gamma)=(1+\Gamma)/(2\sqrt{2})$, as already derived in
\cite{OD}. If one ignores this instability and continues the solutions
of (\ref{eq:sp1},\ref{eq:sp2},\ref{eq:sp3}) to the right of the line
$u=u_{c}(p=2,\Gamma)$ one finds that the memory onset condition
Eq. (\ref{eq:mop}) is fulfilled along the dashed line in the left
panel of Fig. \ref{fig:phasediagram}. As the fixed points are already
unstable in this regime, this line has no direct physical meaning
though, and is depicted only for illustration. We note that for
$\Gamma=1$ both the instability of the fixed point and the onset of
memory occur at the same value of $u=1/\sqrt{2}$. In the statics a de
Almeida-Thouless instability is found at this point and replica
symmetry is broken at smaller values of $u$
\cite{ParisiBiscari}. Interestingly, one does not find any transition
at any positive co-operation pressure $u>0$ for full anti-correlation
$\Gamma=-1$, i.e. in the case of zero-sum games.

To verify the onset of non-ergodicity further we have performed
simulations in which two copies of the system with the same
realization of the disorder are generated and in which the dynamics is
started from two independent sets of random initial conditions drawn
from a flat distribution over $x_i(0)\in[0,2]$. The two resulting
trajectories of the system are labelled by $\{x_i^a(t)\}$ and
$\{x_i^b(t)\}$ respectively and we have measured the squared distance
$d^2=N^{-1}\sum_i (x_i^a(t)-x_i^b(t))^2$ in the stationary
state. Results are depicted in Fig. \ref{fig:p_2_distrough} (left
panel).  For convenience we plot the re-scaled distance $d^2/q$. We
find that $d^2$ is essentially zero above $u_c(\Gamma)$, confirming
the ergodicity and independence of the stationary state on initial
conditions in this regime. Below $u_c$ we find that $d^2>0$,
indicating the existence of multiple stationary states and the
pronounced relevance of initial conditions. For $\Gamma=-1$ no memory effects are observed at any $u>0$ as expected from the theory.

\begin{figure}[t!]
\begin{center}
\includegraphics[width=6.5cm]{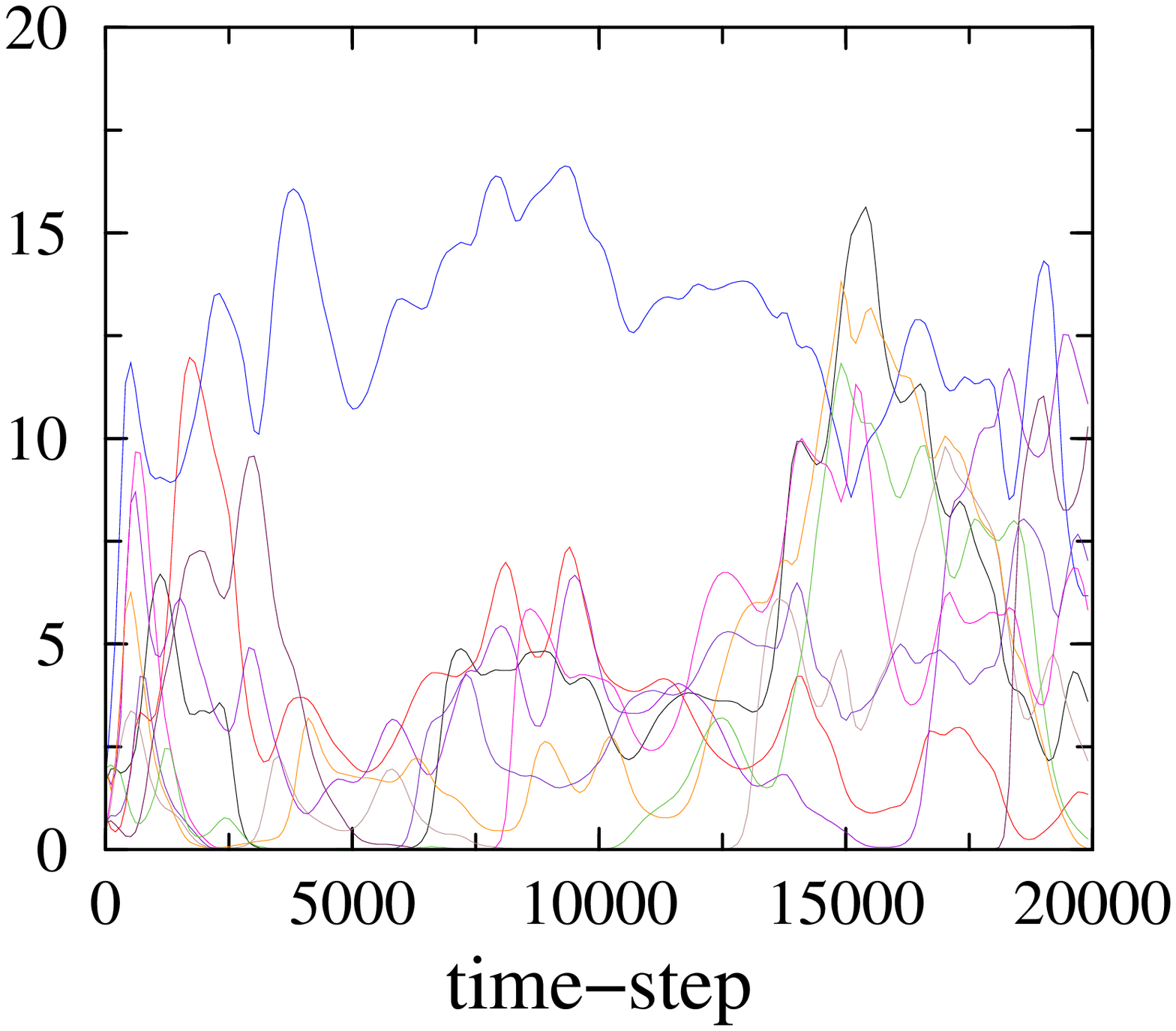}~~~~~
\includegraphics[width=6.5cm]{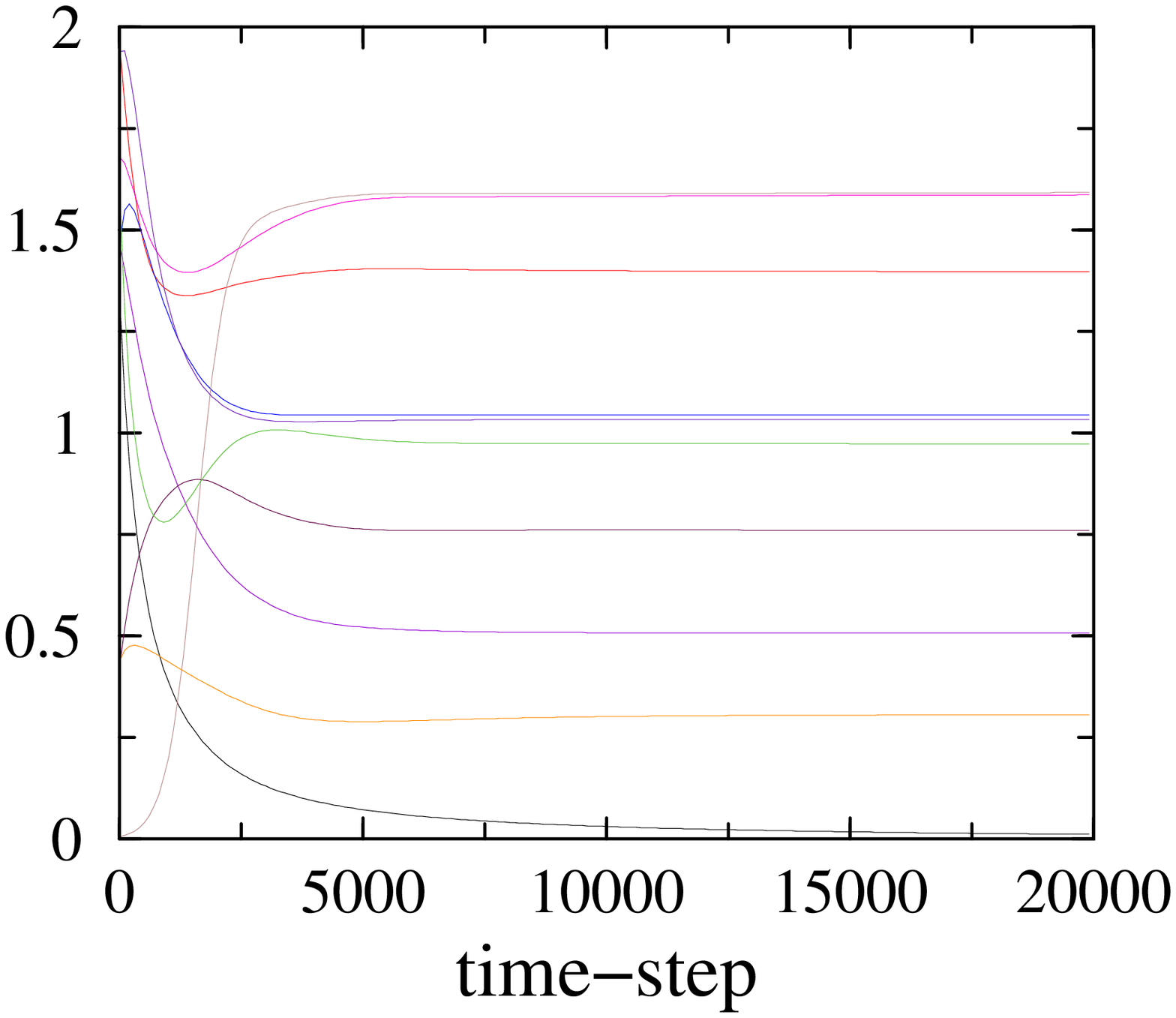}
\caption{\label{fig:p_2_trajectories}Trajectories $\{x_i(t)\}_t$ for $10$ individual species in single runs of simulations of the model with $p=2$ at fully asymmetric couplings, $\Gamma=0$. Left: $u=0.2$ below the transition, where fixed points are unstable; right: $u=1.0$ above the transition. Simulations are for $N=500$ agents runs for $20000$ time steps. }
\end{center}
\end{figure}

We have secondly measured the relative `roughness' $h/q$ of the
trajectories, where $h=N^{-1}\sum_i
(\avg{x_i(t)^2}_t-\avg{x_i(t)}_t^2)$ and $\avg{\dots}_t$ denotes a
time-average in the stationary state. As shown in the right panel of
Fig. \ref{fig:p_2_distrough} all trajectories are flat ($h=0$)
irrespective of $\Gamma$ above $u_c(\Gamma)$, verifying that the
system indeed evolves into a fixed point in this regime. For symmetric
couplings $\Gamma=1$ fixed points are also reached below the
transition. For $-1<\Gamma<1$ we find non-zero values of $h$ below the
transition, so that we conclude that the system does not necessarily
evolve into fixed points here\footnote{The numerical data for $h$
appears to show some dependence on system size, running time and
time-window over which measurements are taken in the regime of
volatile trajectories below $u_c(\Gamma<1)$ (i.e. when $h>0$). The
observations that fixed points are attained for all $\Gamma$ above
$u_c$ and for all $u$ for $\Gamma=1$ are however not affected.  More
extensive simulations seem appropriate in order to check whether the
apparently decreasing values of $h$ as $u$ approaches zero at
$\Gamma<1$ is genuine or a numerical artifact.}. This behaviour is
also confirmed in Fig. \ref{fig:p_2_trajectories} where we show
typical trajectories of the system at $\Gamma=0$ (fully uncorrelated
couplings) below and above and the transition, see also
\cite{OD2} for a similar figure. For $\Gamma=1$ one finds that the system evolves into a fixed point irrespective of $u$ (trajectories not shown here).

We conclude the section concerning the model with $p=2$ by some brief
summary of the effects of asymmetry in the couplings. As discussed
above even partial asymmetry changes the nature of the phase
transition as compared to the fully symmetric case. This was already
pointed out in \cite{OD,OD2}. For symmetric couplings the system
evolves into a fixed point asymptotically both above and below
$u_c(\Gamma=1)=1/\sqrt{2}$. Below the transition fixed points are
however not unique, hence the breaking of ergodicity and dependence on
initial conditions. For $\Gamma<1$ we find that the system reaches a
fixed point asymptotically above $u_c(\Gamma)$, but not necessarily
below, where instead one finds volatile trajectories in which species
can almost die out and then recover to large concentrations at later
times. No transition is observed for $\Gamma=-1$ (full
anti-correlation), and the system here evolves into a unique fixed
point for all $u>0$. The effects of asymmetry on the diversity $1/q$
are demonstrated in Fig. \ref{fig:p_2qphi} (right panel). Generally
speaking asymmetry increases the diversity of species in the
stationary state (equivalently decreasing $\Gamma$ leads to a higher
fraction of surviving species). While this effect is small for large
values of $u$ some significant dependence on $\Gamma$ is observed for
small co-operation pressure $u$.

\begin{figure}[t!]
\begin{center}
\includegraphics[width=6.5cm]{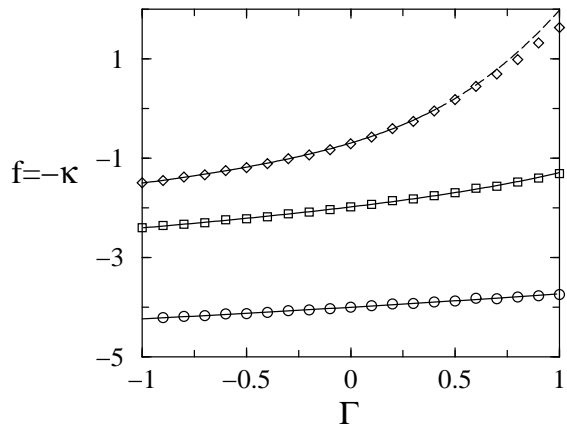}
\caption{\label{fig:fitness_p_2} Average fitness $f=-\kappa$ versus $\Gamma$ for the model with $p=2$. Symbols are from simulations ($u=0.5,1,2$ from top to bottom), simulation parameters as in Fig. \ref{fig:p_2qphi}. The solid lines marks the predictions of the analytical theory, for $u=0.5$ the theoretical line has been continued as dashed line into the phase where the ergodic theory is no longer valid.}
\end{center}
\end{figure}

Finally the effects of asymmetry on the fitness of the species in the
stationary state are demonstrated in Fig. \ref{fig:fitness_p_2}. The
average fitness is here given by $f=N^{-1}\sum_i x_i f_i[\bx]$ with
$f_i[\bx]=-2ux_i-\sum_j J_{ij}x_j$, and it turns out that $f_i\equiv
f$ for all surviving species $i$. From the analytical calculation one
finds $f=-\kappa$. As shown in the figure an increased asymmetry in
the couplings leads to a reduced overall fitness. Again the effect is
strong for small $u$ but negligible for large co-operation pressure
(where the diagonal term proportional to $u$ dominates the
interactions). Thus, in systems with symmetric couplings fewer species
tend to survive than in systems with uncorrelated or anti-correlated
interactions, but these surviving species have higher fitness than the
many survivors of systems with asymmetric or anti-symmetric couplings.

\subsection{Cubic interaction ($p=3)$}
\begin{figure}[t!]
\begin{center}
\includegraphics[width=6.5cm]{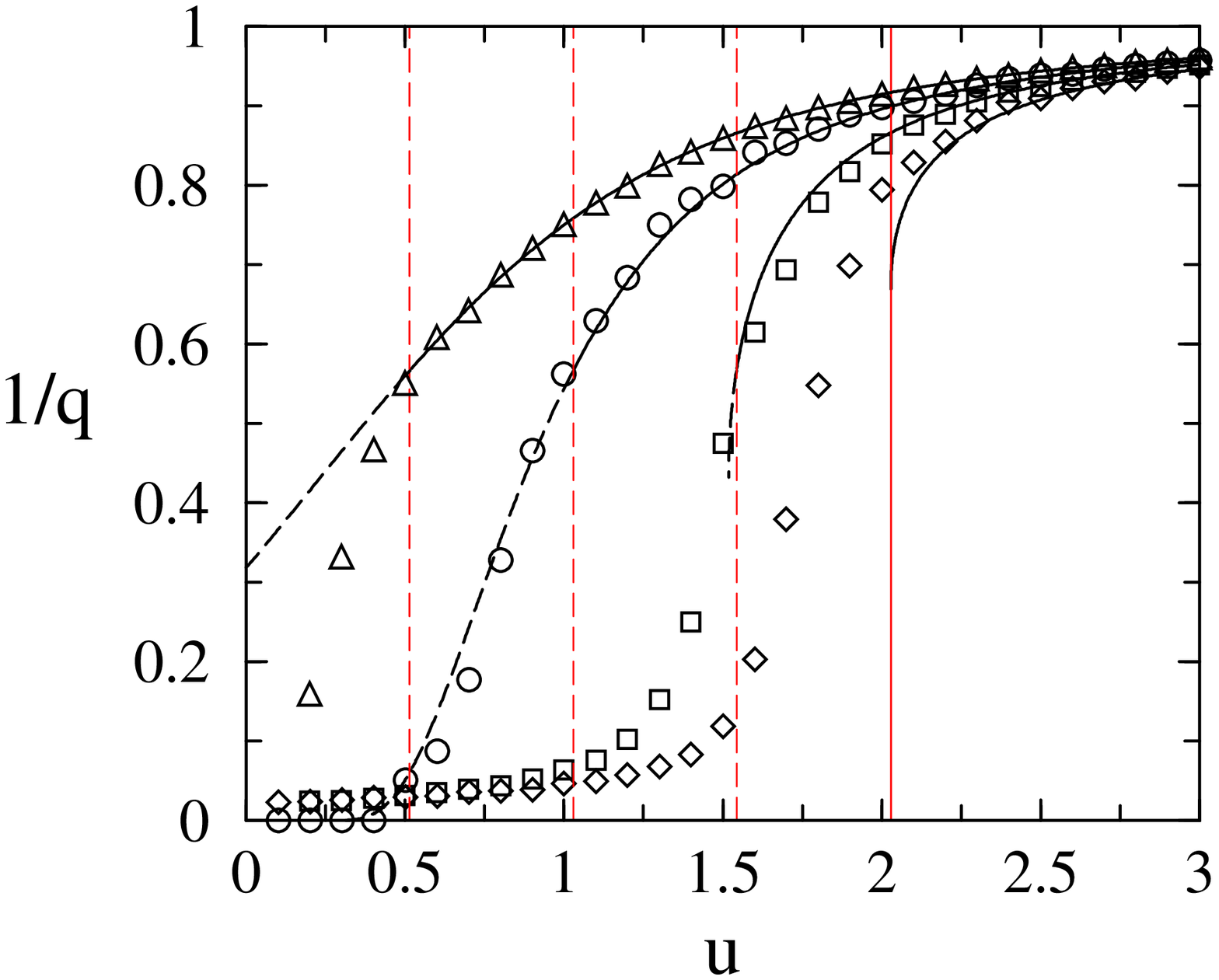}~~~~~
\includegraphics[width=6.5cm]{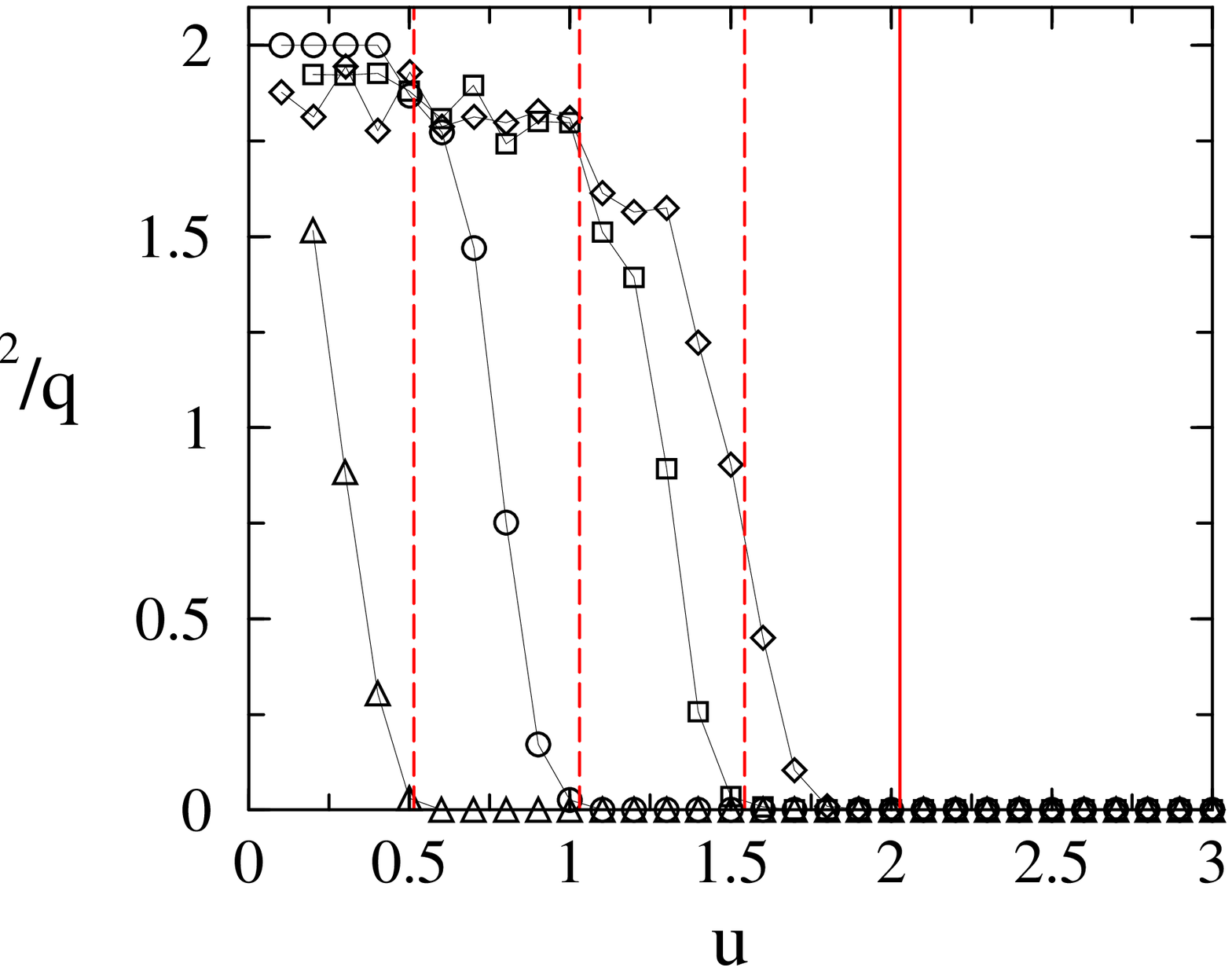}
\caption{\label{fig:p_3a2} Reciprocal order parameter $q$ and distance $d^2/q$ versus $u$ for the model with $p=3$. Symbols are from simulations (triangles: $\Gamma=-0.5$, circles: $\Gamma=0$, squares: $\Gamma=0.5$, diamonds: $\Gamma=1$). Solid lines for $1/q$ are the analytical predictions in the ergodic phase $u>u_c$, continued as dashed lines into the non-ergodic phases. Vertical lines indicate the predicted locations of the transitions. Simulations were performed for $N=200$ species, started from random initial conditions drawn with flat distribution from $x_i(0)\in[0,2]$ and run for $3000$ time-steps. Results are averages over $20$ samples of the disorder.}
\end{center}
\end{figure}

For $p=3$ we find a distinctively different behaviour of the
system. Eqs. (\ref{eq:sp1}, \ref{eq:sp2}, \ref{eq:sp3})
admit solutions only for $u>u^*(\Gamma)$, and below this value no
ergodic fixed point solutions can be found. The obtained value
$u^*(\Gamma=1)\approx 2.028$ has already been reported in the context
of a replica analysis of the model with symmetric couplings
\cite{OF1}. Note also that the limits $\lim_{u\downarrow u^*} q$,
$\lim_{u\downarrow u^*} \chi$ and $\lim_{u\downarrow u^*} \kappa$
exist and take finite values. As indicated in
Fig. \ref{fig:phasediagram} this type of transition is preceded by an
onset of instability of the fixed points (determined by
Eq. (\ref{eq:instp})) for values of $\Gamma<\Gamma_0\approx 0.7$ in
the sense that the instability sets in as $u$ is lowered starting in
the ergodic phase before solutions of the saddle point equations cease
to exist. We note that the onset of instability again occurs along a
line in the $(u,\Gamma)$ plane, with
$u_c(p=3,\Gamma)=u_c(p=3,\Gamma=0)(1+\Gamma)$, where
$u_c(p=3,\Gamma=0)\approx 1.028$\footnote{It turns out that
Eqs. (\ref{eq:sp1},\ref{eq:sp2},\ref{eq:sp3}) have two branches of
solutions above $u^*$, both solutions merge at $u^*$. For
$\Gamma>\Gamma_0$ we find that the fixed points are stable along the
physical branch and that the condition (\ref{eq:instp}) for the onset
of instability is fulfilled at
$u_c(p=3,\Gamma)=u_c(p=3,\Gamma=0)(1+\Gamma)$ only along the
unphysical branch. Thus no instability is observed and a transition of
the above type (iii) is the first one to occur when $u$ is lowered
from infinity. For $\Gamma<\Gamma_0$ the fixed point becomes unstable
along the physical branch and the instability transitions precedes the
failure of solutions to exist.}.  We find that solutions of the
saddle-point equations exist for all $u>0$ if $\Gamma$ is
(sufficiently) negative \footnote{Some numerical inaccuracies are
encountered for $\Gamma\approx 0$ when determining the line along
which solutions cease to exist. The solid line in the phase diagram
for $p=3$ (right panel of Fig. \ref{fig:phasediagram}) could therefore
be investigated more carefully near $\Gamma=0$. At this point
instabilities of the fixed points have already set in and solutions
are unphysical, so that the precise location of the point at which
solutions cease to exist is immaterial for the behaviour of the
model.}.

Results from numerical simulations of the model with $p=3$ are presented in
Fig. \ref{fig:p_3a2}. Given the cubic interaction we limit ourselves
to relatively small system sizes $N=200$ and to moderate running times
here. Results in the ergodic phases are however not affected
significantly by these constraints in computing time. We observe good
agreement between the results for the order parameter $q$ as obtained
from theory and experiment in the ergodic region, with only slight
discrepancies close to the transitions due to finite-size
effects. Again asymmetry in the couplings increases the diversity and
number of surviving species. The right panel of Fig. \ref{fig:p_3a2}
confirms that non-ergodicity sets in at $u_c(\Gamma)$, indicated by
non-zero distances between the stationary states obtained when
starting the dynamics at fixed disorder from different random initial
conditions as explained above.

To summarise we have found a regime
$u>\mbox{max}(u_c(\Gamma),u^*(\Gamma))$ in which our ergodic theory
applies and matches the simulations. For $\Gamma<\Gamma_0$ we then
find an intermediate regime $u_*<u<u_c$ in which solutions of the
saddle point equations exist, but in which the attained fixed points
are unstable and the stationary state depends on initial
conditions. No such intermediate regime appears to be found for
$\Gamma>\Gamma_0$. Finally below $u^*$ no solutions of the equations
for the order parameters in the stationary ergodic state exist. The
transition at $u^*$ is not marked by any divergence in the analytic
solutions for $(q,\chi,\kappa)$ as $u\downarrow u^*$.
\begin{figure}[t!]
\begin{center}
\includegraphics[height=5cm]{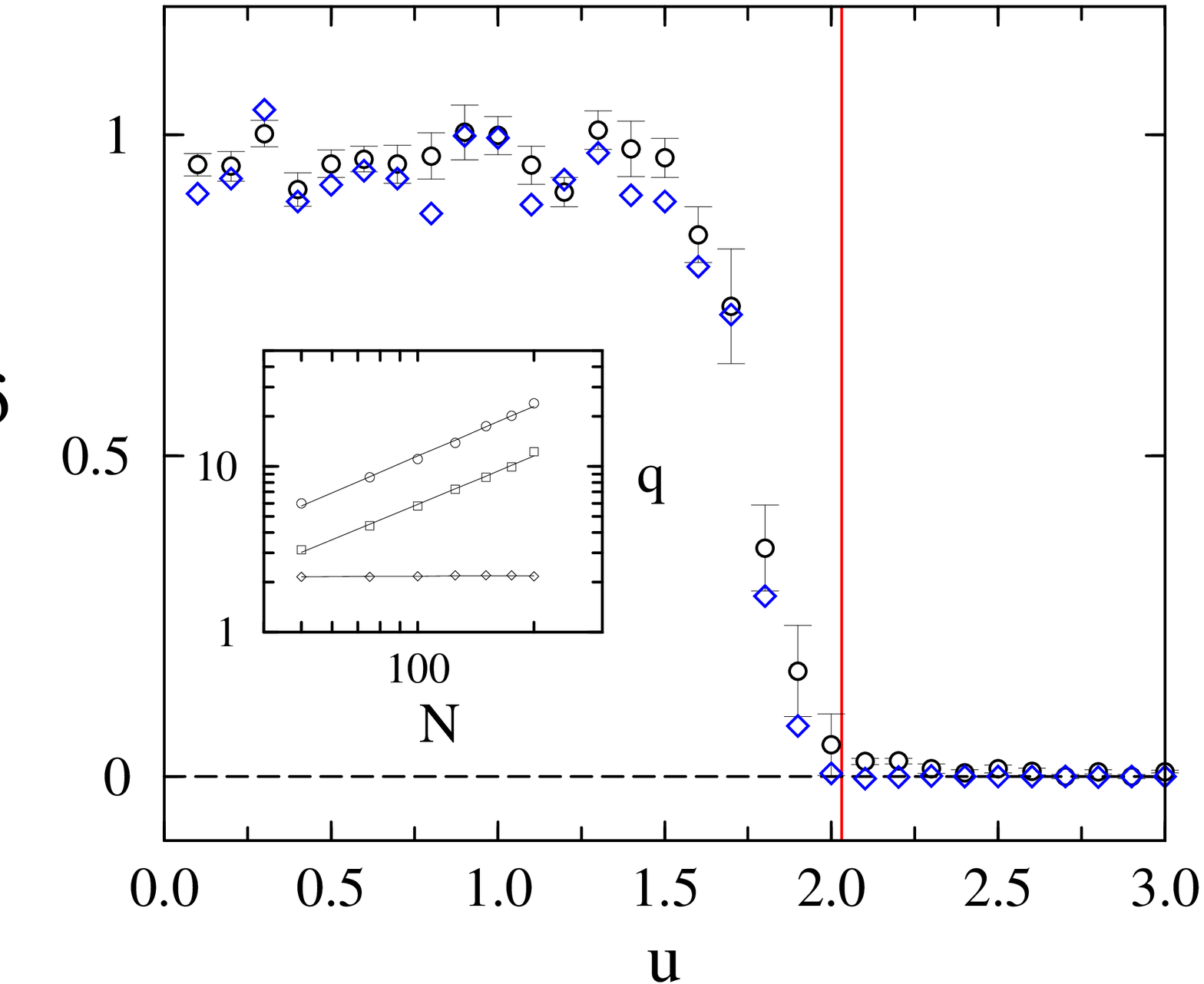}~~~
\includegraphics[height=5cm]{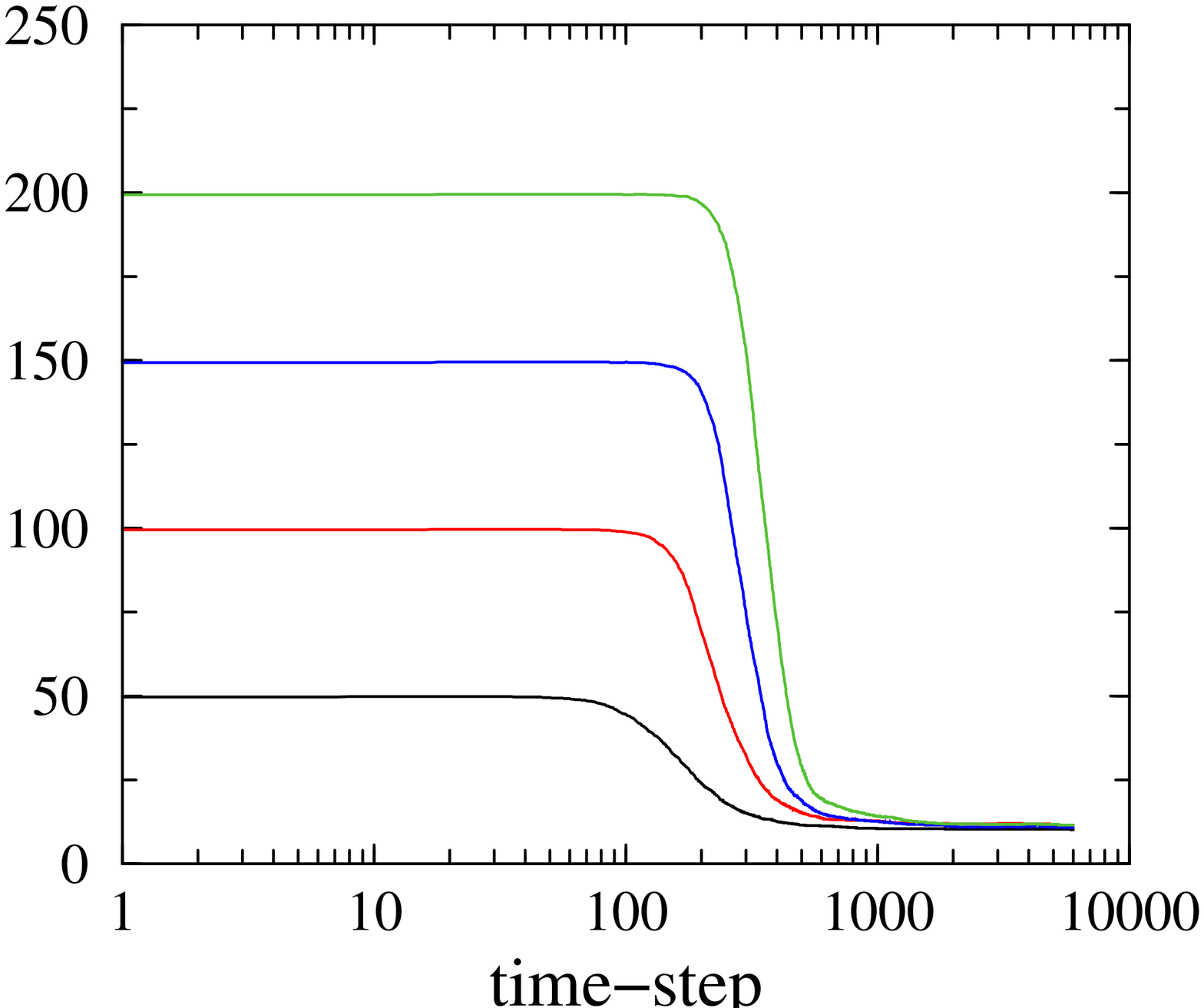}
\caption{\label{fig:exponent} {\bf Left:} Exponents $\gamma$ (circles) and $-\delta$ (diamonds) (see text) as functions of $u$ for the model with $p=3$ ($\Gamma=1$). Each data point is obtained by performing simulations at fixed $u$ for $N=50,75,\dots,200$, simulations are run for $6000$ steps, averages over $10$ samples of the disorder are taken, and subsequent fits to power laws $q\sim N^\gamma$ and $\phi\sim N^\delta$ are performed to obtain $\gamma$ and $\delta$. Vertical bars indicate the standard error of $\gamma$. Vertical solid line marks $u^*(\Gamma=1)$. Inset: Log-log plots of $q$ vs $N$ for $u=1.0$ (circles), $u=1.5$ (squares) and $u=2.5$ (diamonds) and the corresponding fits to power laws. (The data for $q$ at $u=2.5$ in the inset has been multiplied by a factor of two for convenience.) {\bf Right:} Temporal evolution of the number $N_+$ of species with concentration $x_i(t)>0.01$ for the same parameters, and at $u=1$. Different curves are obtained from simulations of different system sizes $N=50,100,150,200$ from bottom to top.}
\end{center}
\end{figure}
To conclude this section we report some indications about the
behaviour of the system below $u^*$ based on numerical simulations (we
restrict to $\Gamma=1$). While the persistent order parameters
converge to finite values in the limit of large $N$ above $u^*$ we
observe strong finite size effects for $u<u^*$. The numerical data are
consistent with power law behaviour $q\sim N^\gamma$ and $\phi\sim
N^\delta$ with $\gamma>0$ and $\delta<0$ below $u^*$, see
Fig. \ref{fig:exponent}. This divergence of $q$ with $N$ seems to
suggest that only a sub-extensive number of species $\phi N\sim
N^{1+\delta}$ survives in this regime ($\delta< 0$) in the
thermodynamic limit. Given the constraint
$\lim_{N\to\infty}N^{-1}\sum_i x_i=1$, we then conclude that these
surviving species each must have concentrations scaling as $x_i\sim
N^{-\delta}$, which in turn results in $q=N^{-1}\sum_i x_i^2\sim
N^{-\delta}$, i.e. we expect $\gamma=-\delta$. Within the accuracy of
our simulations this is indeed what we observe, see
Fig. \ref{fig:exponent}. More precisely our simulations are consistent
with $\gamma=-\delta=1$ at low values of $u$, so that we expect only a
{\em finite} number of species to survive in this regime ($\phi
N\sim{\cal O}(N^0))$, even in the thermodynamic limit. This unusual
effect is indeed confirmed upon plotting the number of species which
are still alive as a function of time for different system sizes, see
the right panel of Fig. \ref{fig:exponent}. Closer
inspection shows that the final number of surviving species in these
simulations all fall into the range $10\leq N\phi\leq 12$, even if the
system size is varied by a factor of four from $N=50$ to $N=200$. A
similar effect was observed in a replicator model with an explicit
extinction threshold
\cite{TokitaYasumoti}.

While our simulations confirm convincingly that $\gamma=-\delta=0$
above $u^*$, and while they seem to suggest that $\gamma=-\delta=1$
sufficiently far below the transition, the accuracy of our numerical
experiments do not allow us to make any statement as to whether this
onset of non-extensivity occurs continuously or discontinuously,
i.e. whether the exponents drop instantly to zero when the transition
at $u=u^*$ is crossed (from below) or whether one observes a
cross-over.

\section{Concluding remarks}
In summary we have demonstrated how generating functionals can be used
to study random replicator models. In the case of symmetric couplings
this approach is able to reproduce the equations describing the
ergodic stationary states obtained from static replica studies, but in
addition the analysis of the dynamics also allows to address
replicator systems with asymmetric interactions and to make accurate
analytical predictions for the order parameters in their fixed point
regimes and the resulting phase diagrams.

Our analysis of Gaussian RRM indicates that the effect of asymmetry
and anti-symmetry in the couplings is to increase the stationary value
of the diversity parameter $1/q$ and with it the number of surviving
species. This is the case for both models studied here, with
interactions between $p=2$ and $p=3$ species respectively. Asymmetry
in the couplings also reduces the fitness of the surviving
species. Furthermore the range of the ergodic phases seem to be
consistently larger the higher the degree of anti-symmetry in the
couplings becomes. The symmetry or otherwise of the interactions also
determines the behaviour of the system in the non-ergodic
phases. Replicators with symmetric couplings evolve into fixed points
in all regions of the phase diagram. In the symmetric case one expects
a large number of such fixed points in the phase below $u_c$
\cite{DO}, and initial conditions determine which of these is reached,
as indicated by the memory-onset. No such phase (and hence no MO
transition) is found in models with asymmetric or anti-symmetric
couplings. Instead, volatile trajectories are here observed below the
transition, and individual species may come close to extinction
$x_i(t)\ll 1$, but then recover to macroscopic values $x_i(t)\sim{\cal
O}(1)$.

Models in which the self-interaction is of a lower order than the
terms coupling distinct species appear to exhibit a phase in which the
fraction of surviving species scales as $\phi\sim 1/N$, so that only a
{\em finite} number of species survives in the long-run even in the
thermodynamic limit. Possibly parallels with condensation effects in
growth models of complex networks with quenched fitnesses of links
and/or nodes can be drawn, as the dynamics of such
networks obeys growth laws which are similar to the
replicator equations studied here \cite{gin1,gin2}.

While we have here concentrated on relatively simple single-population
systems with Gaussian couplings, similar approaches may be taken to
address more intricate models of replicators. These could include
multi-population models of game theory or dynamical models of chemical
metabolic reactions, which progress in proportion to the concentration
of reaction partners and in which stoichiometric coefficients might be
modelled in terms of quenched random couplings. Multi-population
replicator equations of game theory correspond to so-called bi-matrix
games \cite{Book6}, and it would here be interesting to study the relation between
the fixed points of the replicator dynamics and the Nash equilibria of
such games \cite{berg}, also as a function of the degree of anti-correlation in
the payoff matrices and the co-operation pressure. Some work is
currently in progress along these lines.

\section*{Acknowledgements}
This work was supported by the European Community's Human Potential
Programme under contract HPRN-CT-2002-00319, STIPCO. The author would
like to thank G Bianconi, A C C Coolen, M Marsili and D Sherrington
for helpful discussions.

\end{document}